# Rossby wave radiation by an eddy on a beta-plane: Experiments with laboratory altimetry

Yang Zhang and Y. D. Afanasyev[a]

[a]*Memorial University of Newfoundland, St. John's, Canada*

[a]Corresponding author address:

Yakov Afanasyev

Memorial University of Newfoundland, St. John's, NL, Canada

E-mail: afanai@mun.ca

http://www.physics.mun.ca/~yakov

2**Abstract**

Results from the laboratory experiments on the evolution of vortices (eddies) generated in a rotating tank with topographic β-effect are presented. The focus of the experiments is on the far-field flow which contains Rossby waves emitted by travelling vortices. The surface elevation and velocity fields are measured by the Altimetric Imaging Velocimetry. The experiments are supplemented by shallow water numerical simulations as well as a linear theory which describes the Rossby wave radiation by travelling vortices. The cyclonic vortices observed in the experiments travel to the northwest and continuously radiate Rossby waves. Measurements show that initially axisymmetric vortices develop a dipolar component which enables them to perform translational motion. A pattern of alternating zonal jets to the west of the vortex is created by Rossby waves with approximately zonal crests. Energy spectra of the flows in the wavenumber space indicate that a wavenumber similar to that introduced by Rhines for turbulent flows on the β-plane can be introduced here. The wavenumber is based on the translational speed of a vortex rather than on the root-mean-square velocity of a turbulent flow. The comparison between the experiments and numerical simulations demonstrates that evolving vortices also emit inertial waves. While these essentially three-dimensional non-hydrostatic waves can be observed in the altimetric data they are not accounted for in the shallow water simulations.

## I.  INTRODUCTION

Meso-scale vortices (eddies) are an essential element of the dynamics of the turbulent oceans. They provide a strongest signal in the snapshots of the circulation measured by the satellite altimetry. Eddies are intimately linked to narrow zonal flows (jets) observed in midlatitudes (Maximenko et al. [1, 2]). Altimetric signal due to zonal jets is much more subtle than that of eddies. For this reason the jets are called latent in the oceans. Although it is known that the existence of the zonal jets is due to the β-effect, the details of their generation are still a subject of ongoing discussion. One of the mechanisms discussed is related to the radiation of Rossby waves by eddies. The radiation of Rossby waves results in the creation of gyres (β-plumes) elongated in the zonal direction. Each β-plume consists of two jets flowing in the opposite directions (Davey and Killworth[3], Rhines[4] and Afanasyev et al.[5]). β-plumes can be described within the framework of linear dynamics. This implies that linear modes can provide a significant control of the entire flow which include eddies, jets and Rossby waves. A more general question is then to what extent the linear dynamics is important. In their recent numerical and theoretical studies Berloff and Kamenkovich[6] analyzed linear modes in the idealized ocean circulation containing multiple jets and eddies and showed that certain properties of (generally nonlinear) eddies can be understood in terms of the linear modes. The linear modes themselves can be modified by the background flow. An analysis of an ocean gyre circulation in a spectral space performed recently by Chen et al.[7] showed that the zonal jets (also called striations) can be interpreted either within linear dynamics context as almost zero-frequency Rossby waves or as a result of the nonlinear eddy propagation.

The present study is motivated by the oceanographic phenomena mentioned above. In what follows, we consider an idealized setup where Rossby waves are radiated by single vortices propagating on the topographic polar β-plane in the rotating tank. We focus on the far-field, away from the immediate vicinity of the vortex, where the properties of the flow were not measured in detail in previous experimental studies. In particular, we identify the horizontal wave patterns and measure the spectral characteristics of the flow. We supplement our laboratory experiments with numerical simulations and theory. The results of the experiments are used as a "benchmark" to compare with the analytical results and the results of the simulations. The comparison between linear theory and nonlinear simulations or experiment allows us to determine to what extent the linear dynamics can predict the pattern of waves in a generally nonlinear flow.

Although a number of theoretical and numerical studies addressed the radiation of Rossby waves by vortices, laboratory investigations are relatively few. This is perhaps due to the fact that it is difficult to observe the wave field using traditional laboratory techniques. Visualization with dye allows one to record the vortex trajectory and observe its evolution but gives little information about the wave field. Particle Image Velocimetry (PIV) technique is a well-known experimental tool and can be used to measure velocity within the vortex and in its vicinity. It is, however, more challenging to measure the far-field velocity in a large tank. Most of the laboratory studies were focused on investigating the dynamics of the vortex motion on the β-plane (e.g. Carnevale, Kloosterziel and van Heijst[8], van Heijst[9], Stegner and Zeitlin[10], Zavala Sanson and van Heijst[11], Flor and Eames[12]). Flor and Eames[12] used PIV to measure the velocity profiles in monopolar cyclonic vortices created by suction or stirring. They also studied the trajectories of the vortices and compared the measured trajectories with those predicted by a theoretical mechanistic model based on integral relations for the Rossby force and a lift force. In this study we use a recently developed experimental technique, Altimetric Imaging Velocimetry (AIV) to overcome the shortcomings of the earlier techniques and to observe and measure the flow field both within the vortex and in the entire tank with the same spatial and temporal resolution. The AIV technique measures the gradient of the surface elevation field. Integration of the gradient in the horizontal plane allows one to obtain the surface elevation field which can also be interpreted as the pressure field at the surface. The surface elevation is one of the major dynamic fields that can be used to describe the Rossby waves. The measured gradient can be further used to obtain the velocity field in the flow using quasi-geostrophic equations.

Radiation of Rossby waves by stationary perturbations was discussed in application to different oceanographic problems including the flows induced by localized buoyancy sources (Stommel[13], Joyce and Speer[14], Davey and Killworth[3], Helfrich and Speer[15]). A similar solution for an atmospheric tropical cyclone was given by Chan and Williams[16]. A linear radiation process is easy to understand. When a perturbation is steady, long Rossby waves with frequency approaching zero have nearly zonal wavecrests. As a result of the radiation, the perturbation "stretches" to the west forming a ridge/trough. According to geostrophy, two zonal jets form along the slopes of the ridge/trough. The formation of the β-plume circulation due to the source of buoyant fluid was illustrated in the laboratory experiments on the polar (quadratic) β-plane by Afanasyev et al.[5].

Vortices in a turbulent flow are not at rest; they propagate due to interaction with other vortices or with the background flow. Single vortices on the β-plane propagate due to nonlinear effect of the induced wave field on the vortex. Cyclonic vortices travel to the north-west while anticyclonic vortices travel to south-west while radiating the Rossby waves. Thus, it is important to include the source motion into the radiation problem. Lighthill[17] gave a general solution of the linear problem for a travelling transient forcing. The external forcing can be due to a wind-stress curl present in a finite region of space, travelling with constant velocity and varying in magnitude over some time period. A particular case is a steady forcing for which the forcing frequency is zero. Lighthill's solution gives the frequency of the Rossby waves as a function of wavenumber in *x*- and *y*-directions. In the case of the steady forcing the solution predicts the relation between the *x*- and *y*-wavenumbers (a curve in wavenumber space) for any particular velocity of the forcing.

In Sec. II of this paper, we describe the setup of the laboratory apparatus as well as the altimetry technique. In Sec. III the setup of the shallow water numerical model is described. Theory of the Rossby wave radiation by a travelling vortex is presented in Sec. IV. In Sec. V the results of the laboratory experiments, numerical simulations and theory are reported. Concluding remarks are given in Sec. VI.

## II. LABORATORY SETUP AND TECHNIQUE

In our laboratory experiments, a cylindrical tank of radius $R = 55$ cm was filled with water of depth $H_0 = 10$ cm (Figure 1). The tank was installed on a rotating table and was rotated anticlockwise at a constant angular rate $\Omega = 2.4$ rad/s.

The vortices in the tank were generated by suction of fluid from below the surface. A thin tube connected to a pump was placed on the bottom of the tank such that the opening of the tube was directed upward (Figure 1). The suction creates a localized sink which manifests itself as a depression on the surface. Water converging to the sink creates a cyclonic vortex in the presence of the background rotation. The vorticity in the core of the vortex is created by stretching the background vorticity. Both the suction rate (determined by the voltage, *V*, applied to the pump) and the duration of suction, $\Delta t$, were varied in the experiments such that vortices of different strength and size were created. The control parameters for five experiments are given in Table 1. Table 1 also summarizes the main characteristics of the vortices measured right after the forcing period. The characteristics include the surface elevation, $\eta_v$, in the center of each vortex, the maximum azimuthal velocity, $U_{\theta v}$, the vortex radius, $R_v$, the total kinetic energy, *K*, the mean Rossby number, $\text{Ro}_v = U_{\theta v}/(f_0 R_v)$, and the ratio of the vortex azimuthal velocity to its translational velocity, $A = U_{\theta v}/U_t$. Here $f_0 = 2\Omega$ is the Coriolis parameter.

The surface of water in the tank when in a state of a solid-body rotation is a paraboloid, such that the depth of water is given by

$$h(r) = H_0 + \frac{\Omega^2}{2g}\left(r^2 - \frac{R^2}{2}\right), \qquad (1)$$

where $r$ is the distance to the axis of rotation (center of the tank) and *g* is the gravitational acceleration. A dynamical effect of the radial variation of depth is similar to that due to a variation of the Coriolis parameter on a rotating planet. The dynamical equivalence of these effects follows from the conservation of potential vorticity (PV). The center of the tank corresponds to the North pole of the planet. Due to the quadratic variation of the depth of the layer, the laboratory system corresponds to a so-called polar β-plane (or γ-plane) such that the Coriolis parameter is $f = f_0 - \gamma r^2$, where $f_0 = 2\Omega$ and $\gamma = \Omega^3/gH_0$ (e.g. Afanasyev and Wells[18]). In this study, however, we also use a regular β-plane approximation for comparison with numerical simulations and theory. The β-plane where the Coriolis parameter varies linearly in the South-North direction can be defined with respect to a reference distance from the pole $r_0$. A local Cartesian coordinate is introduced at this reference "latitude" such that the *x* and *y* axes are directed to the east and the north respectively. The β-parameter is then defined as

$$\beta = 2r_0 \Omega^3 / (gh(r_0)) \qquad (2)$$

In the experiments, the vortices were created at $r_0 = 30$ cm where $\beta = 0.1$ cm$^{-1}$s$^{-2}$.

Altimetry method[19] was used to measure two components of the gradient $\nabla \eta = (\partial \eta / \partial x, \partial \eta / \partial y)$ of the perturbation surface elevation $\eta$ in the horizontal plane (*x*, *y*). The $\nabla \eta$ field was measured with a spatial resolution of approximately 2 vectors per millimeter which translates into the array of size 2300×2300, with a temporal resolution of 5 fields per second. The surface velocity of the flow is then determined using quasigeostrophic approximation:

$$\mathbf{U} = \frac{g}{f_0}(\mathbf{n} \times \nabla \eta) - \frac{g}{f_0^2}\frac{\partial}{\partial t}\nabla \eta - \frac{g^2}{f_0^3} J(\eta, \nabla \eta), \qquad (3)$$

where **U** is the horizontal velocity vector and **n** is the vertical unit vector. Note that while we measure an "exact" (within experimental accuracy) pressure gradient, $\nabla p = \rho g \nabla \eta$, at the surface, the velocity field is determined more accurately when the flow is closer to being quasigeostrophic. According to the Taylor-Proudman theorem, in a rapidly rotating flow, the surface velocity is a good approximation for the velocity in the entire column of water except the Ekman layer at the bottom. Table 1 shows that the Rossby number, $\text{Ro}_v = 2U_{\theta v}/f_0 R_v$, that characterizes the relative vorticity in the core of the vortex exceeds unity in experiments 1 – 4 (immediately after the forcing stops). Here $U_{\theta v}$ is the maximum velocity of the vortex and $R_v$ is the radius of the maximum velocity. In these experiments where $\text{Ro}_v > 1$, the validity of the quasigeostrophic approximation was not satisfied within the vortex cores in the initial period of their evolution. As a consequence, the velocity within the cores calculated with Eq. (3) differed from the "real" velocity. Later in these experiments as flow decayed, the values of the Rossby number dropped below unity. In the experiment 5 the validity of the quasigeostrophic approximation was satisfied at all times. The flow beyond the cores of the vortices (which is the primary focus of this study) was always well within the bounds of the approximation in all experiments such that the velocity given by the AIV was quite accurate.

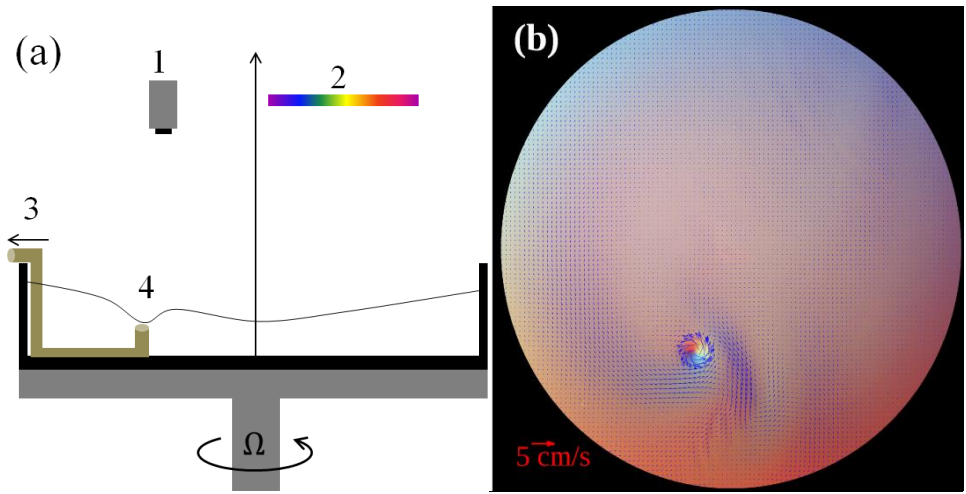

Figure 1. (Color online) (a) Sketch of the experimental setup (left): digital camera (1), high brightness display showing the color mask (2), the fluid in the tank is pumped out through a thin tube (3) to generate a cyclonic vortex (4). (b) Top view of the flow (color) with a superposed velocity field (vectors) measured by the AIV at 19 s in experiment 1.

Table 1. Experimental parameters

| Experiment | $\Delta t$ (s) | V (Volt) | $\eta_v$ (cm) | $R_v$ (cm) | $U_{\theta v}$ (cm/s) | K (cm$^2$/s$^2$) | Ro$_v = \frac{2U_{\theta v}}{f_0 R_v}$ | $A = \frac{U_{\theta v}}{U_t}$ |
|---|---|---|---|---|---|---|---|---|
| 1 | 7.7 | 7.9 | -0.19 | 1.7 | 5.8 | 7.0 | 1.4 | 6.1 |
| 2 | 2.6 | 7.9 | -0.16 | 0.90 | 5.8 | 2.7 | 2.6 | 9.7 |
| 3 | 3.2 | 6.9 | -0.14 | 0.91 | 5.7 | 2.3 | 2.6 | 11 |
| 4 | 8.4 | 6.0 | -0.19 | 1.1 | 6.2 | 6.0 | 2.3 | 7.8 |
| 5 | 7.1 | 9.0 | -0.20 | 2.7 | 4.4 | 8.2 | 0.67 | 3.8 |

### III. Numerical model

We consider vortices on the β-plane in a shallow water model,

$$\frac{\partial}{\partial t}(\nabla^2 - k_d^2)\psi = -\beta\psi_x - J(\psi, \nabla^2\psi) - \nu\nabla^6\psi - \lambda\nabla^2\psi \qquad (4)$$

where $\psi$ is the stream function and $k_d = f_0/\sqrt{gH_0} = 0.05\ cm^{-1}$ is the reciprocal of the deformation radius. The bi-harmonic diffusion term $\nu\nabla^6\psi$ is routinely used in the simulations of two-dimensional turbulence in order to effectively remove motions at the smallest scales (e.g. Bracco et al.[20], Maltrud and Vallis[21]). The term $\lambda\nabla^2\psi$ represents linear Ekman bottom friction. A particular value of the friction coefficient, $\lambda = 0.03\ s^{-1}$, was chosen to model the flow decay in our laboratory experiments. The spatial differencing was implemented using a spectral method which implies the periodic boundary conditions in both directions. For the sake of numerical efficiency and stability, we used semi-implicit time scheme "AB3CN" by Boyd[22]. The Jacobian operator was discretized using a third-order Adams–Bashforth scheme, the second-order Crank-Nicholson scheme was applied to the linear part including $\beta$ term and damping terms. The numerical domain was set to be a square of 110 cm wide with 512 grid points along each side. The value of the $\beta$ parameter, $\beta = 0.1\ cm^{-1}s^{-2}$ in the simulations was the same as that in the experiments. The simulations were performed in a rectangular (double periodic) domain on a regular β-plane, rather than in a circular geometry and a polar β-plane (as in our experiments) for the purpose of easier comparison and interpretation of the results since in the majority of previous theoretical or numerical studies a regular β-plane was used.

IV. **Theory**

Suppose we have a turbulent flow on an f-plane (e.g. Afanasyev and Craig 2013[23]). Energy distribution between motions of different spatial scale is established by nonlinear (triad) interactions. The energy spectrum is isotropic in wavenumber space. The flow on the f-plane is similar to a purely two-dimensional turbulent flow in non-rotating fluid except perhaps for the presence of the Ekman layer at the bottom. Let us now suddenly "switch on" the β-effect. While the flow still remains balanced on small scales where nonlinear terms prevail over the β-effect, it will be unbalanced on larger scales where a quasi-geostrophic (QG) type balance is required. In what follows we look for an additional component of the flow that is required to balance the initially specified turbulent flow. We consider the scales starting from the scale where the turbulent flow starts to "feel" the β-effect (the Rhines scale) which corresponds to the scale of the largest vortices/eddies formed in the turbulent cascade, and larger. Zonal structures are formed in physical space to satisfy the quasi-geostrophic type balance; meanwhile in spectral space the energy distribution becomes anisotropic because energy can now cascade directly to the zonal modes such that the additional component of the flow will be in the form of Rossby waves. At large scales it is sufficient to consider a linear dynamics at least as first approximation, The turbulent vortices (assumed known) constitute a forcing in the wave equation. In what follows, for simplicity, we consider only one vortex.

Let us follow a well-known derivation of a QG equation in order to obtain the forcing terms. Consider a flow on a β-plane ($x$, $y$) where the Coriolis parameter varies linearly in $y$-direction as

$$f = f_0 + \beta y \ . \tag{5}$$

The flow consists of an assumed known eddy component with velocity $\mathbf{U} = (U_x, U_y)$ and the additional component due to the β-effect (the Rossby wave component) with velocity $\mathbf{u} = (u_x, u_y)$. Both components can be related to pressure fields, expressed in terms of surface elevations $\eta_0$ and $\eta$ respectively, via geostrophic relations

$$\mathbf{U} = \frac{g}{f_0} \mathbf{k} \times \nabla \eta_0 \ , \quad \text{and} \quad \mathbf{u} = \frac{g}{f_0} \mathbf{k} \times \nabla \eta \ , \tag{6}$$

where $\mathbf{k}$ is the vertical unit vector and $g$ is the acceleration due to gravity. Potential vorticity of the flow can defined in a usual manner

$$q = \frac{f + \zeta_0 + \zeta}{h} \ , \tag{7}$$

where relative vorticities are $\zeta_0 = \frac{g}{f_0} \nabla^2 \eta_0$ and $\zeta = \frac{g}{f_0} \nabla^2 \eta$. The fluid depth is given by

$$h = H - h_b + \eta_0 + \eta \ , \tag{8}$$

where $H$ is the mean depth and $h_b$ is the bottom height. Assuming that $h_b$, $h_0$ and $h$ are small compared to $H$ and $\zeta_0$ and $\zeta$ are small compared to $f$ we rewrite $q$ in the form

$$q = \zeta_0 + \zeta + \beta y - \frac{f_0}{H}(\eta + \eta_0 - h_b) \ . \tag{9}$$

The dynamics of the flow is governed by the conservation equation for the potential vorticity

$$\left(\frac{\partial}{\partial t}+(\mathbf{U}+\mathbf{u})\cdot\nabla\right)q=0, \tag{10}$$

which, after some algebra, can be rewritten in the form

$$\frac{\partial}{\partial t}\left(\nabla^2-k_d^2\right)(\eta_0+\eta)+(\mathbf{U}+\mathbf{u})\cdot\nabla\nabla^2(\eta_0+\eta)+k_d^2(\mathbf{U}+\mathbf{u})\cdot\nabla h_b+\beta\frac{\partial}{\partial x}(\eta_0+\eta)=0, \tag{11}$$

where $k_d$ is the reciprocal of the radius of deformation, $k_d=R_d^{-1}=f_0/\sqrt{gH}$. To derive Eq. (11) we used the identity $(\mathbf{U}+\mathbf{u})\cdot\nabla(\eta_0+\eta)=0$ which results from the geostrophic relations (6). In our further analysis we consider a domain with a flat bottom such that the term containing $\nabla h_b$ is equal to zero. Note that this term gives a vertical velocity due to the flow over topography which can result in interesting effects in a class of problems where the bottom topography is important (Spall[24], Spall and Pickart[25]). We can neglect the quadratic term $(\mathbf{u}\cdot\nabla)\nabla^2\eta$ in Eq. (11) assuming that the wave field is relatively weak. The term $(\mathbf{U}\cdot\nabla)\nabla^2\eta_0$ describes the advection of the relative vorticity of a vortex by its velocity field. It can be shown that this term vanishes if we assume that the vortex is axisymmetric. Indeed, in that case the advection is just a rotation of a vorticity distribution given by $\nabla^2\eta_0$ around the center of the vortex. The translation of a vortex is determined by the wave velocity which is significant inside the vortex. Note that the vortex velocity field generates large difference in the Coriolis force between the northern and southern parts of the eddy. The mean Coriolis force drives the (cyclonic) vortex to the North. The advection of the vortex can be approximated by a constant translation velocity $\mathbf{U}_t$ such that $(\mathbf{u}\cdot\nabla)\nabla^2\eta_0\approx(\mathbf{U}_t\cdot\nabla)\nabla^2\eta_0$. Outside the vortex, the term $(\mathbf{u}\cdot\nabla)\nabla^2\eta_0$ is negligible since both $\mathbf{u}$ and $\eta_0$ are relatively weak. Linearized Eq. (11) becomes:

$$\frac{\partial}{\partial t}\left(\nabla^2 - k_d^2\right)\eta + \beta\frac{\partial}{\partial x}\eta = -\frac{\partial}{\partial t}\left(\nabla^2 - k_d^2\right)\eta_0 - \beta\frac{\partial}{\partial x}\eta_0 - (\mathbf{U}_t \cdot \nabla)\nabla^2\eta_0 \ . \tag{12}$$

In the rotating systems Ekman dissipation can be important. This effect can be easily included in a form of the Ekman pumping. Vertical velocity at the boundary of the bottom Ekman layer is proportional to the relative vorticity of the flow

$$w_E = \frac{1}{2}\delta_E(\zeta_0 + \zeta) \ , \tag{13}$$

where $\delta_E = (2\nu/f_0)^{1/2}$ is the thickness of the Ekman layer and $\nu$ is the kinematic viscosity of fluid. Introducing the Ekman number, $E = 2\nu/f_0 H^2$, we obtain

$$w_E = \frac{1}{2}E^{1/2}k_d^{-2}f_0\nabla^2(\eta_0 + \eta) \ . \tag{14}$$

The part of $w_E$ which is due to the relative vorticity of the vortex, will serve as an additional forcing term in Eq. (12) while the part due the relative vorticity of the wave will act as a damping term. With these additional Ekman terms, Eq. (12) becomes

$$\frac{\partial}{\partial t}\left(\nabla^2 - k_d^2\right)\eta + \beta\frac{\partial}{\partial x}\eta + \lambda\nabla^2\eta = -\frac{\partial}{\partial t}\left(\nabla^2 - k_d^2\right)\eta_0 - \beta\frac{\partial}{\partial x}\eta_0 - (\mathbf{U}_t \cdot \nabla)\nabla^2\eta_0 - \lambda\nabla^2\eta_0 ,$$

$$\tag{15}$$

where $\lambda = E^{1/2}f_0/2$ is the Ekman coefficient.

Assuming the RHS of Eq. (15) is known, we transform the equation into Fourier space

$$\frac{\partial\tilde{\eta}}{\partial t} + (i\omega + \omega_E)\tilde{\eta} = -\frac{\tilde{F}}{k^2 + k_d^2} \ . \tag{16}$$

16Here

$$\omega = \frac{-k_x \beta}{k^2 + k_d^2} \quad , \quad \omega_E = \frac{\lambda k^2}{k^2 + k_d^2} \tag{17}$$

are the Rossby wave frequency and the Ekman frequency respectively, $F$ is the RHS of Eq. (15), $\mathbf{k} = (k_x, k_y)$ is the wavevector and tilde denotes the spatial Fourier transform. Solving the first-order differential equation (16) with an initial condition $\tilde{\eta}(t=0) = 0$, we obtain

$$\tilde{\eta} = \exp(-i\omega t - \omega_E t) \int_0^t \frac{-\tilde{F}}{k^2 + k_d^2} \exp(i\omega t + \omega_E t) dt \quad , \tag{18}$$

Let us specify now a particular form of forcing in order to gain further insight into the dynamics of the flow. Consider a vortex traveling with velocity $\mathbf{U}_t$ without changing its spatial structure such that its surface elevation is $\eta_0 = \eta_0(\mathbf{r}')$, where $\mathbf{r}' = \mathbf{r} - \mathbf{r}_0 - \mathbf{U}_t t$ and $\mathbf{r}_0$ is the initial position of the eddy. In the Fourier space this translates into $\exp(-i\mathbf{k} \cdot \mathbf{r}_0 - i\mathbf{k} \cdot \mathbf{U}_t t)\tilde{\eta}_0$ according to the shift theorem. The forcing can then be written as

$$\tilde{F} = \left(-ik_d^2(\mathbf{k} \cdot \mathbf{U}_t) - i\beta k_x + \lambda k^2\right) \exp(-i(\mathbf{k} \cdot \mathbf{r}_0) - i(\mathbf{k} \cdot \mathbf{U}_t)t)\tilde{\eta}_0 \quad . \tag{19}$$

Substituting (19) into (18), we obtain the solution in the form

$$\tilde{\eta} = \frac{i\omega_v - i\omega - \omega_E}{-i\mathbf{k} \cdot \mathbf{U}_t + i\omega + \omega_E} \exp(-i\mathbf{k} \cdot \mathbf{r}_0) \left[\exp(-i\mathbf{k} \cdot \mathbf{U}_t t) - \exp(-i\omega t - \omega_E t)\right] \tilde{\eta}_0 \quad , \tag{20}$$

where we introduced a vortex frequency

$$\omega_v = \frac{k_d^2(\mathbf{k} \cdot \mathbf{U}_t)}{k^2 + k_d^2} \quad . \tag{21}$$

In what follows we use a simple expression for the vortex in the form:

$$\eta_0 = \eta_v M(r)\left[1 + \frac{f_0}{g\eta_v}(U_{ty}x - U_{tx}y)\right], \tag{22}$$

where $M(r)$ is the monopolar component of the vortex which describes its radial structure. The second term in the brackets gives an additional dipolar component which is necessary for the translation of the vortex. The x- and y-coordinates in Eq. (22) are defined in a coordinate system with the origin in the center of the vortex. The radial distribution $M(r)$ can be obtained from the experiments. The ratio of the magnitude of the monopolar to dipolar components is determined by the ratio of the azimuthal velocity of the vortex to its translational velocity.

$$A = \frac{U_{\theta v}}{U_t} = \frac{g\eta_v}{f_0 U_t}. \tag{23}$$

## V. RESULTS

We performed five experiments with different forcing (Table 1); in experiments 2 and 3 the forcing was applied for a relatively short time, $\Delta t = 2 - 3$ s, while in the rest of the experiments the forcing time was $\Delta t = 7 - 8$ s. As a result the vortices in experiments 2 and 3 are of relatively small radius, low amplitude of the surface elevation and, consequently, low energy. The maximum velocity, $U_{\theta v}$, of these low-energy vortices is still relatively high as are the values of parameter $A$ which can be considered as a measure of the nonlinearity. Despite the differences between the control parameters in the experiments, the flows in all of the experiments were qualitatively similar to each other. Fig. 2 shows a typical evolution of the flow.

The color shows $\nabla\eta$ field as recorded by the video camera while the arrows show the velocity field obtained in the post-processing of the color images as described in Sec. II. A local coordinate system with x-axis directed to the East and y-axis directed to the North (the center of the tank is the North pole) is shown in Fig. 2 a. A cyclonic vortex indicated by a circular rainbow-like color pattern, is formed by the forcing (Fig. 2 a, b) and then propagates to the northwest. The vortex radiates Rossby waves; the longer and more zonal waves are to the west of the vortex while shorter waves are trailing behind the vortex, to the East. Note, that the altimetric signal due to the waves are weaker than that due to the vortex. A (global) pattern of the Rossby wave in the entire tank can be easily identified in the distribution of the surface elevation (Fig. 3). The depression of the surface (shown by darker shading) extends westward from the vortex forming a typical β-plume.

Averaging the azimuthal velocity $v_\theta$ and the surface elevation $\eta$ in the azimuthal direction around a vortex center, we obtained the radial profiles of these two quantities. They are shown in the first and second column respectively in Fig. 4 for all five experiments. The total kinetic energy, K, of a vortex can be obtained by integrating $v_\theta^2/2$ over an area which surrounds the moving vortex at each time (here we used a circular area of approximately 5 cm radius). The third column in Fig. 4 gives K as a function of time. The end of the forcing period in each experiment is marked by a cross. During the forcing period, the energy grows linearly, while after the forcing is switched off the energy decays approximately exponentially. By tracking the vortex center (surface elevation minimum), the trajectory of each vortex can be determined. The trajectories for all experiments are shown in the fourth column of Fig. 4. Crosses denote the position of a vortex when the forcing is switched off. Here the x-component of the vortex displacement was measured in the zonal direction while the y-component was measured along the local North (radial) direction with respect to a reference distance from the pole $r_0$. The trajectories are approximately straight lines such that the direction of propagation is approximately at an angle $\alpha = 140^0$. Note that similar, almost straight trajectories were predicted theoretically by Reznik and Dewar[26] (see their Figs. 1 and 2) for vortices with Gaussian or hurricane-like profiles of vorticity.

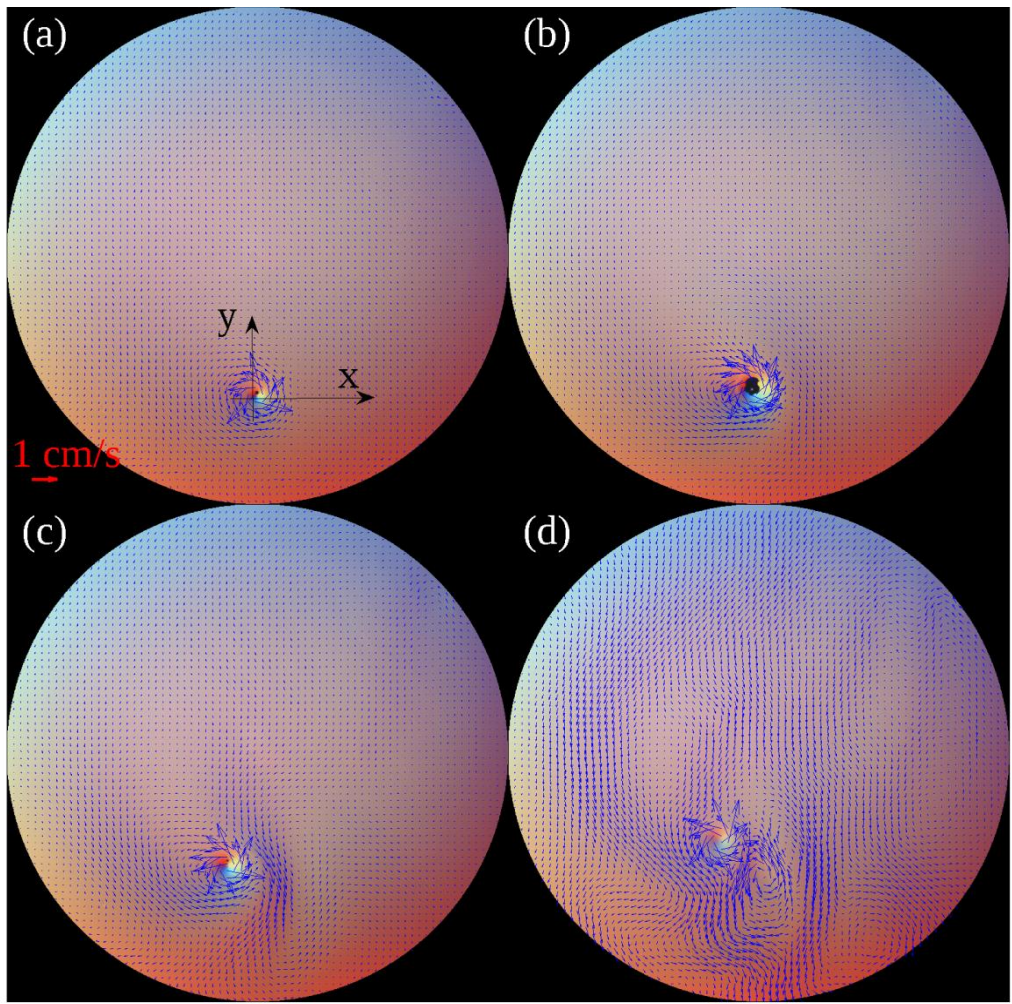

Figure 2(color online). Flow evolution in experiment 1 visualized by AIV: (a) the beginning of forcing, $t = 4$ s; (b) the end of the forcing period when the vortex achieved its maximum strength, $t = 9$ s; (c) and (d) unforced vortex, $t = 16$ s and $t = 28$ s. The blue arrows indicate the velocity field.

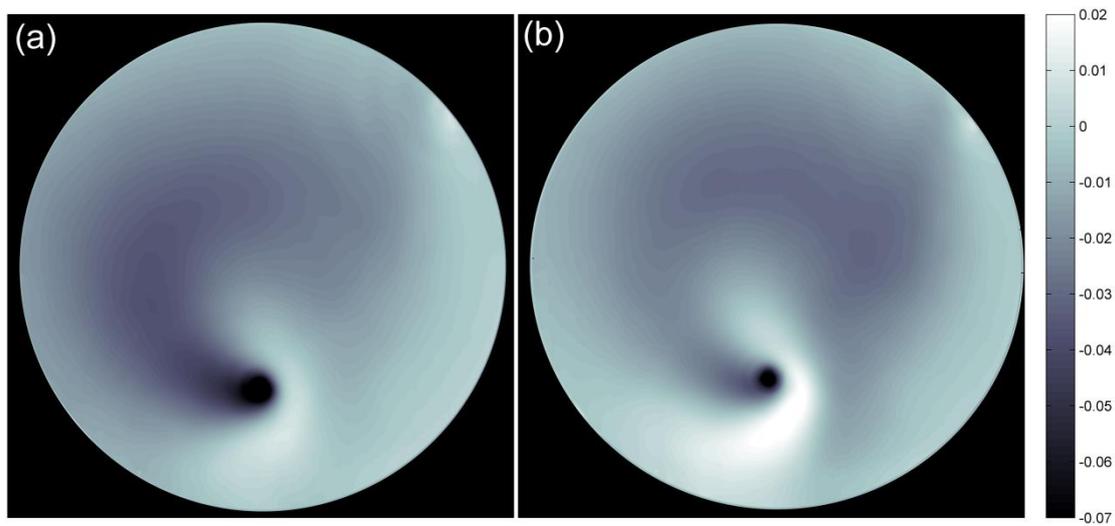

Figure 3. Surface elevation field in experiment 1 at $t = 10$ s (a) and $t = 12$ s (b). Greyscale shows $\eta$ in cm.

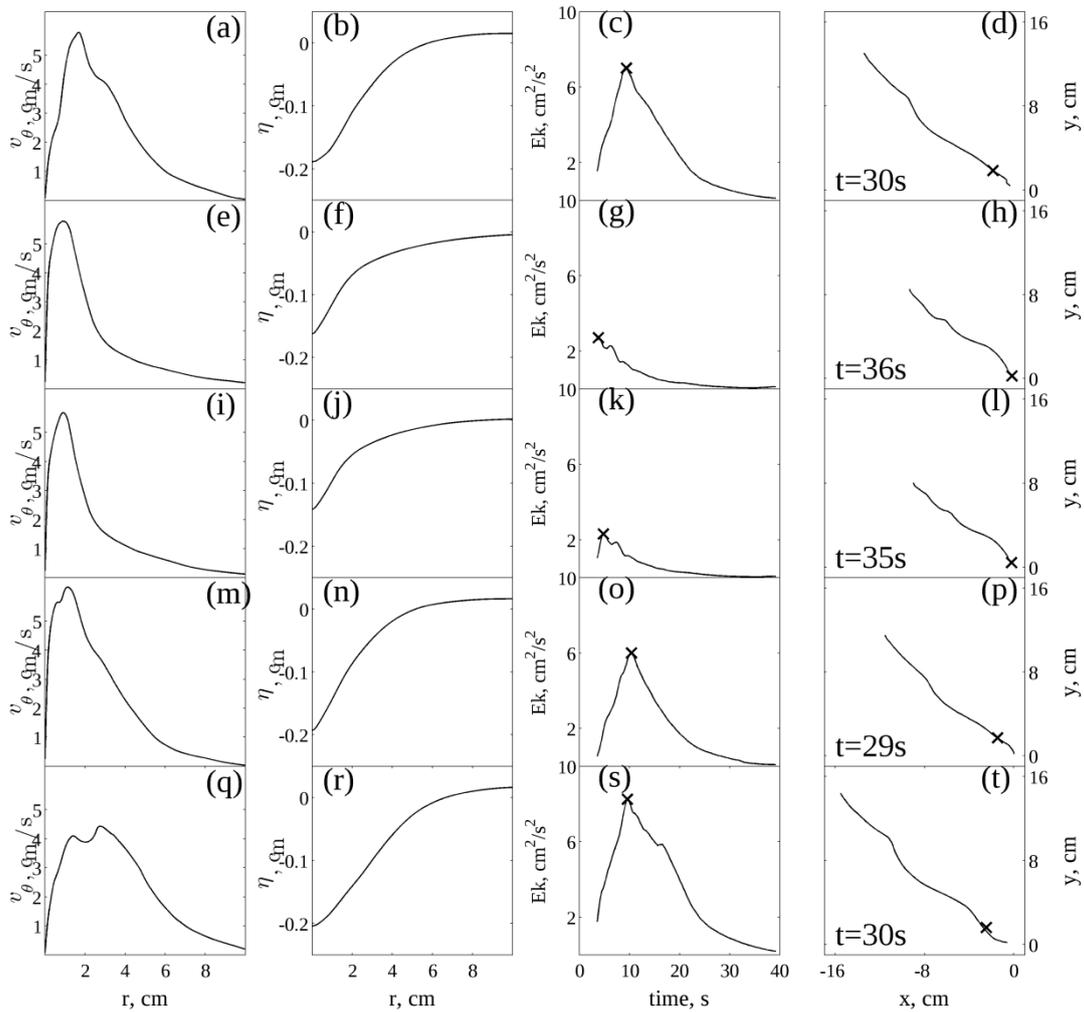

Figure 4. Characteristics of vortices measured in experiments 1 – 5: the azimuthal velocity in cm/s (the first column); the surface elevation in cm (the second column); total kinetic energy as a function of time (the third column) and the vortex trajectory (the fourth column). The total time of travel after the forcing stops is indicated in the last column.

An axially symmetric monopolar vortex cannot perform translational motion unless a dipolar component is added in order to match the velocity inside the vortex to that of translational motion. On the β-plane, the breaking of the axial symmetry of the flow is provided by the β-effect. A so-called β-gyre is formed within the vortex (Reznik[27]). The formation of the dipolar β-gyre is easy to understand. A cyclonic vortex advects water parcels to the North at its eastern side and to the South at its western side. According to the conservation of potential vorticity, the parcels advected to the North acquire anticyclonic relative vorticity to compensate for the increased background vorticity while the parcels advected to the South acquire cyclonic relative vorticity. Thus, the additional vorticity forms a dipole with its axis directed to the North which indicates the primary direction of the translational motion. However, the dipole is also affected by the monopolar velocity field and its axis rotates cyclonically. As a result of this complex nonlinear interaction the axis of the dipolar component and, hence, the direction of the translational motion of the entire vortex is to the northwest. Similar arguments show that anticyclone propagates to the southwest.

We can use the measured fields to reveal the dipolar component of the flow using Fourier transform in the azimuthal direction. In a local polar coordinate system attached to a vortex, the surface elevation field could be decomposed into angular modes:

$$\eta(r,\theta) = M(r) + a(r)\cos\theta + b(r)\sin\theta, \tag{24}$$

where $M(r)$ is the monopolar component which can be calculated as the azimuthal average of $\eta$. The dipolar component is a sum of two orthogonal terms of magnitude $a(r)$ and $b(r)$ respectively. The relative strengths of the two dipolar terms determine the direction of propagation of the vortex. Fig. 5 shows the surface elevation fields due to the monopolar and dipolar components together with the geostrophic velocity fields corresponding to these components. The fields were measured right after the forcing was stopped in each experiment. The dipole (the second column in Fig. 4) has a cyclonic vortex at the southwest and an anticyclone at the northeast such that the axis of the dipole and the main flow induced by the dipole are directed to the northwest as theory predicts. The magnitudes of $a(r)$ and $b(r)$ are relatively small compared to the monopolar term $M(r)$ (the third column in Fig. 5). To confirm that the dipole provides the translation of the entire vortex structure we compared the velocity of the vortex measured in the experiments with that due to the dipole (the fourth column in Fig. 5). The zonal ($x$-) and meridional ($y$-) components of the vortex translational velocity, $\mathbf{U}_t$, were measured by tracking the position of the center of the vortex (minimum $\eta$) and then differentiating with respect to time. The velocity due to the dipole was obtained by averaging the geostrophic velocity

$$\mathbf{u}_{dipole} = \frac{g}{f_0}\mathbf{k} \times \nabla \eta_{dipole}, \qquad (25)$$

where $\eta_{dipole} = a(r)\cos\theta + b(r)\sin\theta$, over the area of the vortex. The comparison between the directly measured velocity $\mathbf{U}_t$, and that calculated from the dipole surface elevation field shows a close match that confirms that the vortex is indeed driven by its dipolar component.

Fig. 6 shows the monopolar and dipolar components at different times in the experiment 1. The sequence in the first column clearly shows that the monopole decays with time. The dipolar fields in the second column in Fig. 6, exhibit an interesting periodic behavior. The dipole inside the vortex is swirled by the monopolar velocity field such that the dipole can even reverse its direction at the center of the vortex (Fig. 6 k and Fig. 6 r). Note that the dipole outside of the vortex remains consistently to the northwest. The swirling of the dipole in the center is due to a nonlinear interaction between the monopolar and dipolar components. As a result of the swirling, the translational motion of the entire vortex structure is oscillatory which can be seen clearly in the time sequences of the translational velocity (the fourth column in Fig. 5). The time of the dipole reverses correlates well with the time when the entire vortex slows down.

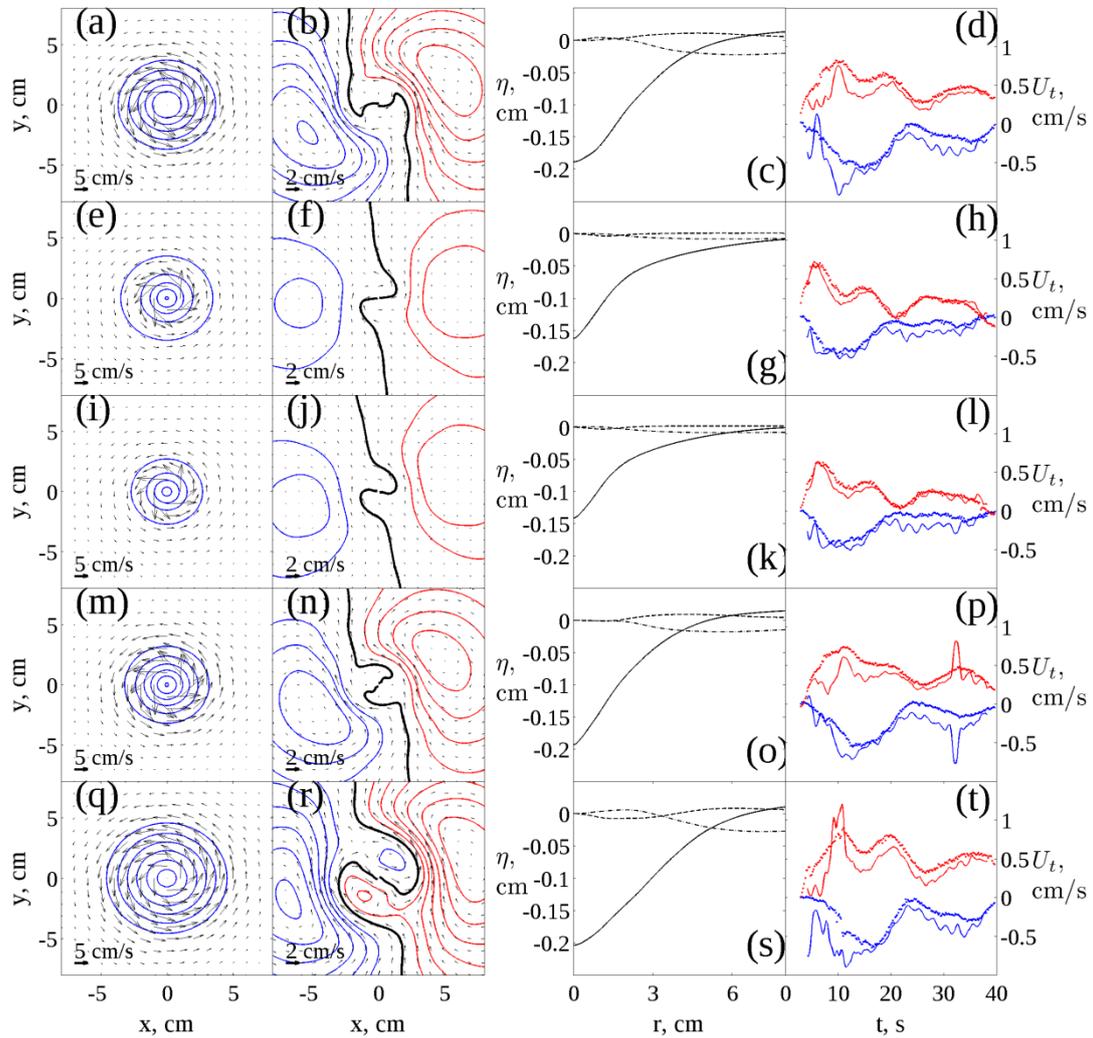

Figure 5. Monopolar and dipolar components of the flow in experiments 1- 5: (the first column) the monopolar component $M$ (contours show $\eta$ in the range from -0.25 cm to -0.03 cm with 0.03 cm interval, arrows show the geostrophic velocity); (the second column) the dipolar component (contours show $\eta$ in the range from -0.2 cm to 0.2 cm with 0.004 cm interval, red lines indicate positive values, the blue lines indicate negative values, the black shows zero $\eta$, arrows show the geostrophic velocity); (the third column) radial profiles $M(r)$ (solid line), $a(r)$ (dashed line) and $b(r)$ (dashed-dot line); (the fourth column) zonal and meridional components of the translation velocity of the vortex measured in the experiments (solid blue and red lines respectively) and the velocity derived from the dipolar component (dotted lines).

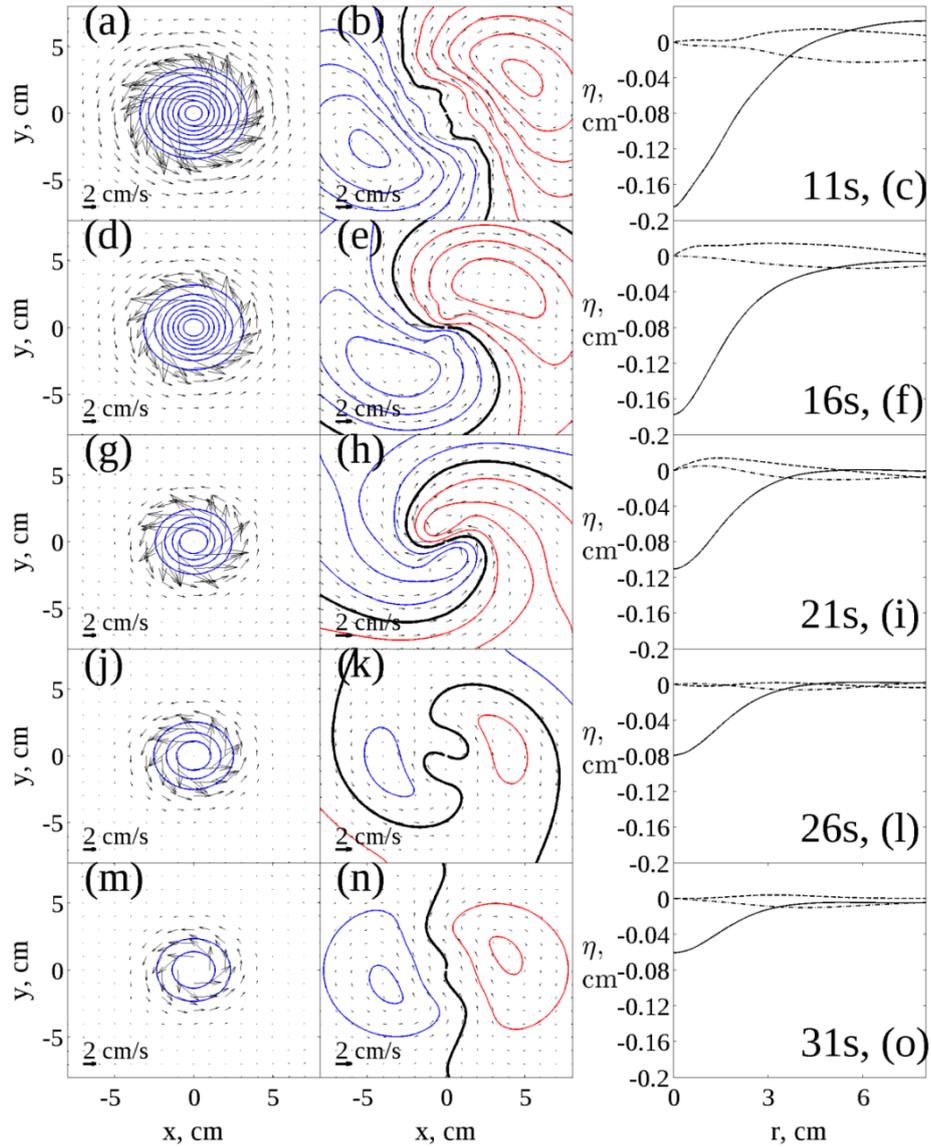

Figure 6. Evolution of the monopolar (the first column) and the dipolar (the second column) components of the flow in experiment 1 at t = 11s, 16 s, 21s, 26 s, 31s. The third column gives *M(r)* (solid line), *a(r)* (dashed line) and *b(r)* (dashed-dot line). The contour lines show $\eta$ with 0.01 cm interval for the monopole and with 0.005 cm interval for the dipole; red/blue line indicates positive/negative values respectively, the black line shows zero $\eta$ and arrows show the geostrophic velocity.

We performed numerical simulations of the flow with the control parameters similar to those in the laboratory experiments (Sec. III). The numerical simulations were initialized with an axisymmetric distribution of the surface elevation mimicking that in the experiment right after the vortex is fully formed. Since the simulations were performed in a rectangular (rather than circular) domain and on a regular (rather than polar) β-plane certain differences between the experimental and numerically simulated flows can be noted. These differences arise from a different geometry and different boundary conditions (such as the presence of a wall in the tank) but they are not crucial for the physical interpretation of the flows.

In order to see to what extent the observed laboratory flows can be explained by the linear theory, here we also present the theoretical solutions together with the experimental and numerically simulated flows. We used vortex profiles obtained in the experiments to specify a translating vortex in the RHS of Eq. (15). The vortex was in a form given by Eq. (22) where we used experimental data to specify the profile of the monopolar component, $M(r)$, and the translation velocity $\mathbf{U}_t$. The field of surface elevation is then given by the inverse Fourier transform of solution given by Eq. (20).

Fig. 7 shows the comparison of the surface elevation fields between the experiment, numerical simulation and linear theory while Fig. 8 shows the relative vorticity. For easier comparison the laboratory fields were interpolated into a local Cartesian coordinate system with its origin fixed at the position of forcing, and $x$- and $y$- axes directed to the East and to the North respectively. Figs. 7 and 8 show the snapshots of the flow at five different times. There are general similarities between the laboratory, numerical and theoretical fields. The vortex travels to the northwest, leaving behind a wave trail. The waves crests have approximately parabolic shape; the waves propagating to the East are short while the waves propagating to the West are long and approximately zonal as one can expect. The vortex decays in magnitude due to the Ekman friction at the bottom as well as due to regular friction in the bulk of the fluid layer. The comparison between numerical simulations and linear theory shows that the theory predicts quite well the pattern of the waves in the far-field. This confirms the approximately linear character of the radiation in spite of the fact that the vortex itself is strongly nonlinear. It is not entirely surprising since we account for the nonlinearity by specifying the translational motion of the vortex.

The differences between the experimental and simulated or theoretical flows are also worth noting. In particular, the perturbations of $\eta$ at the northern part of the domain appear to be propagating much farther westward in the experiments compared to that in the numerical simulations. Most likely the reason is geometric, due to the fact that the tank is circular such that the size of the domain in the x-direction becomes smaller when approaching the center of the tank. As a result, a circumpolar circulation can be easily established there. Similar effect can be important in real atmospheric flows (and, perhaps, to lesser extent in oceanic flows) and is not accounted for in the regular β-plane setup.

Relative vorticity fields in Fig. 8 allow us to see fine features of the flow. The surface elevation (as in Fig. 7) is obtained by integration of the measured $\nabla \eta$ and, as a result, all small-scale features are smoothened. Vorticity, on the other hand, is the results of differentiation which reveals the fine features (the downside of differentiation of experimental data is, of course, that it amplifies noise). A couple of interesting features can be observed in Fig. 8. Firstly, the cyclonic vortex generated by suction wraps the negative (anticyclonic) vorticity around itself and thus becomes partially isolated. There is also evidence of instability in the ring of the anticyclonic vorticity when two small satellite anticyclones form and the cyclonic core of the vortex becomes elliptic. Secondly, inertial waves can be observed in the flow. They only present in the experimental flow and have an appearance of thin filaments within the patches of vorticity around the vortex (indicated by arrows in Fig. 8).

The emission of inertial waves by a travelling barotropic vortex is of interest in the oceanographic context because it provides a path for the energy transfer from mesoscale eddies to motions of smaller scales (submesoscale). Inertial waves should not be confused with inertial oscillations which are inertia-gravity waves (IGW) in the limit when their frequency approaching the Coriolis frequency $f_0$. IGW are surface waves of frequency above $f_0$. Near inertial IGW or inertial oscillations are sometimes called in short "inertial waves" in oceanographic literature. However, inertial waves have frequency below $f_0$ and are three-dimensional waves that can propagate in the bulk of the fluid. They are otherwise known as Kelvin waves[28] or gyroscopic waves. Inertial waves constitute a basis of linear dynamics of rotating fluid (e.g. Greenspan[29]); Rossby waves can in fact be considered as simply a special type of inertial waves (Phillips[30]). Inertial waves are important in the process of adjustment of the flow and can also be regarded as "spontaneously" emitted by an otherwise balanced flow. Some altimetric observations of inertial waves emitted by a meandering coastal flow in a rotating fluid were previously presented by Afanasyev et al.[31]. Since this phenomenon is rarely observed in the experiments it is worth investigating in more detail. Note that here we report on inertial waves of relatively high frequency (although still below $f_0$) compared to the frequency of Rossby waves.

Inertial waves can be easily identified in a sequence of consecutive images of the flow (Fig. 9) or in a video (not shown here) by their curious feature when the phase of the wave propagates toward the source of the wave rather than away from it (as gravity waves do when say a stone thrown in a pond disturbs the surface of water). To visualize the evolution of the waves and to measure their general characteristics a Hovmoeller (space-time) diagram was rendered. The diagram in Fig. 10 shows the distribution of the geostrophic velocity along a straight line at different times. The line was drawn along the vortex trajectory; the velocity component perpendicular to the line was recorded. The vortex, where the velocity changes from positive to negative and is of large magnitude, is visible as white and black bands in the middle of the diagram. The vortex detaches itself from the sink when the forcing stops at approximately 9 s and then moves along the line. The slope of the bands indicates that it moves with an approximately constant velocity. Inertial waves manifest themselves as thin bands above (in front of) and below (behind) the vortex. The inertial waves are superposed on Rossby waves which are of larger scale. The slopes of the bands allows us to measure the phase speed, $c \approx 0.33$ cm/s, while the distance between the lines gives the wavelength, $\lambda \approx 2.7$ cm (which gives the horizontal wavenumber, $k \approx 2.3$ cm$^{-1}$). It is interesting to check the measured properties against the dispersion relation for inertial waves. Figure 8 in Afanasyev et al.[31] shows dimensionless frequency, $\omega/f_0$ as a function of dimensionless wavenumber $kR_d$ for different vertical modes. Here $R_d = (gH_0)^{1/2}/f_0$ is the barotropic radius of deformation. In our case, the waves are of low frequency, $\omega/f_0 \approx 0.17$ and of high wavenumber, $kR_d \approx 50$. The dispersion relation plot in Afanasyev et al.[31] then shows that these particular values of frequency and wavenumber correspond to the vertical mode of the lowest order which has a simplest vertical structure. Although it is difficult to pinpoint the exact mechanism of emission here, we can hypothesize that the emission is of the "spontaneous" type as that described in the theoretical study by Ford et al.[32] This terminology emphasizes that this emission occurs due to the dynamics of the quasi-balanced flow rather than due to an imbalance in the initial conditions. Indeed, our vortex together with its Rossby wave field is approximately balanced within the quasi-geostrophic framework. The (relatively weak) emission of the inertial waves occurs during the entire time of the evolution of the vortex long after the forcing ended. This indicates that this emission is the result of the higher-order dynamics beyond the quasi-geostrophy.

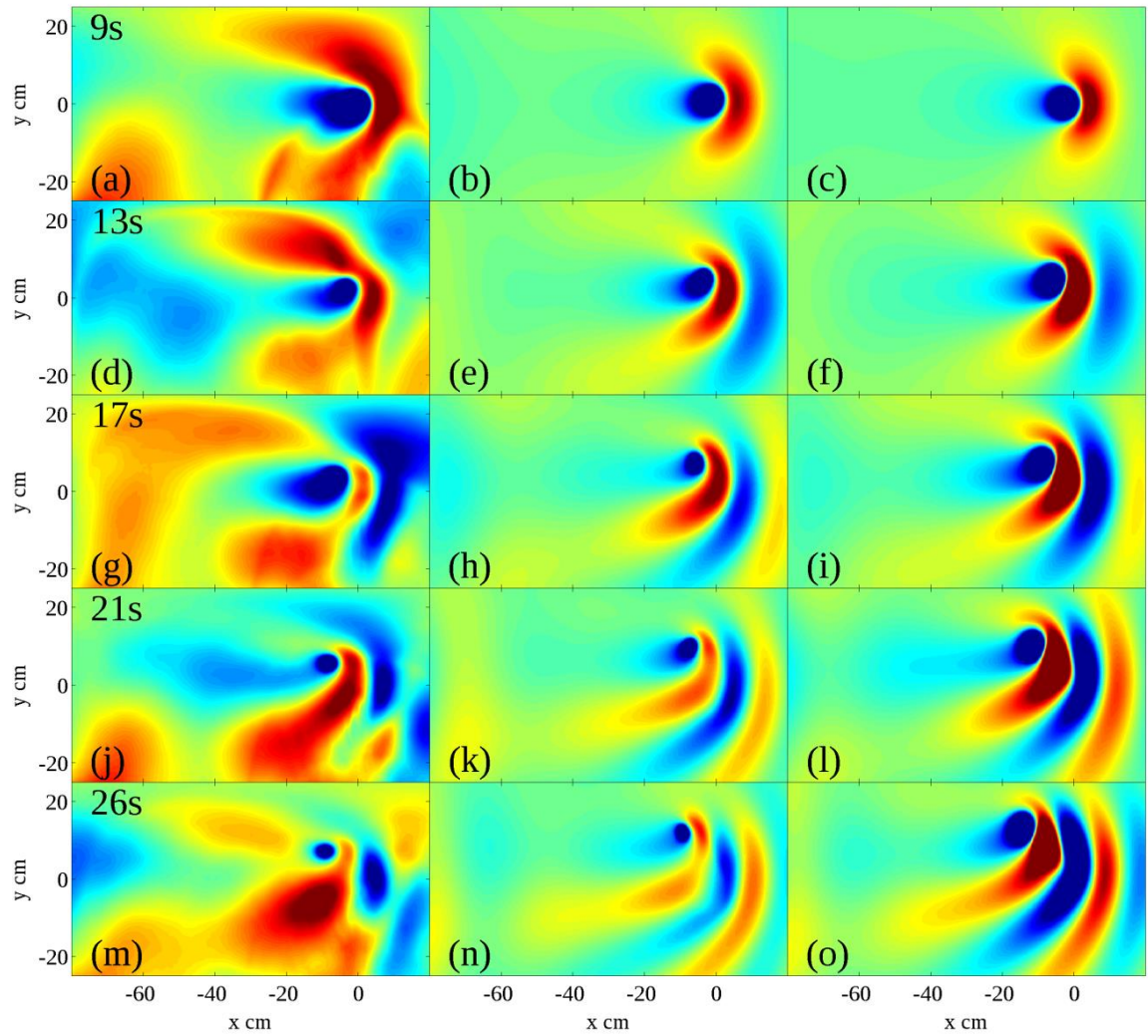

Figure 7. Comparison of the surface elevation fields $\eta$ in the experiment 1 (the first column), numerical shallow-water simulation (the second column) and linear theory (the third column). Values of $\eta$ is in the range between -0.03 cm (blue) to 0.03 cm (red).

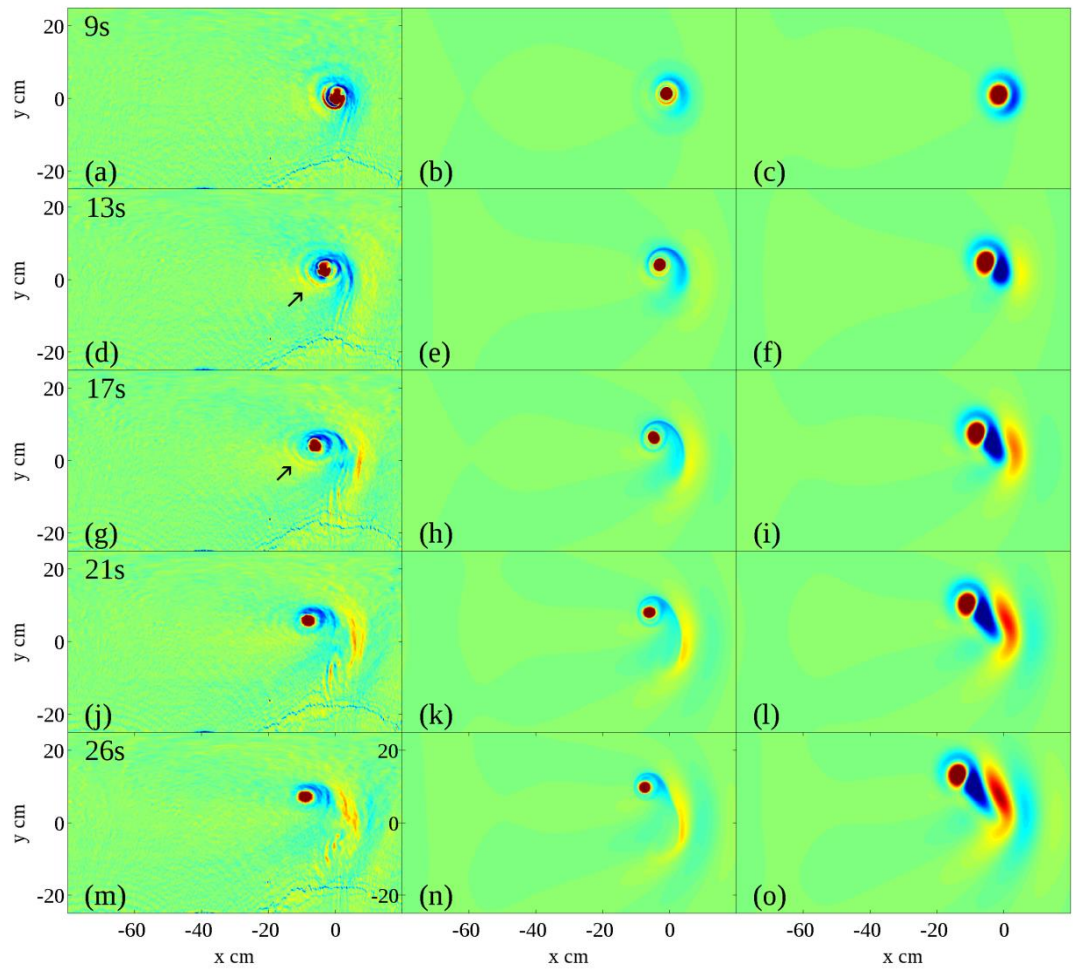

Figure 8. Same as in Fig. 7 but for the relative vorticity fields. Vorticity is normalized by the Coriolis parameter and varies in the range between -0.5 (blue) and 0.5 (red). Arrows indicate the crests of inertial waves emitted by the evolving vortex.

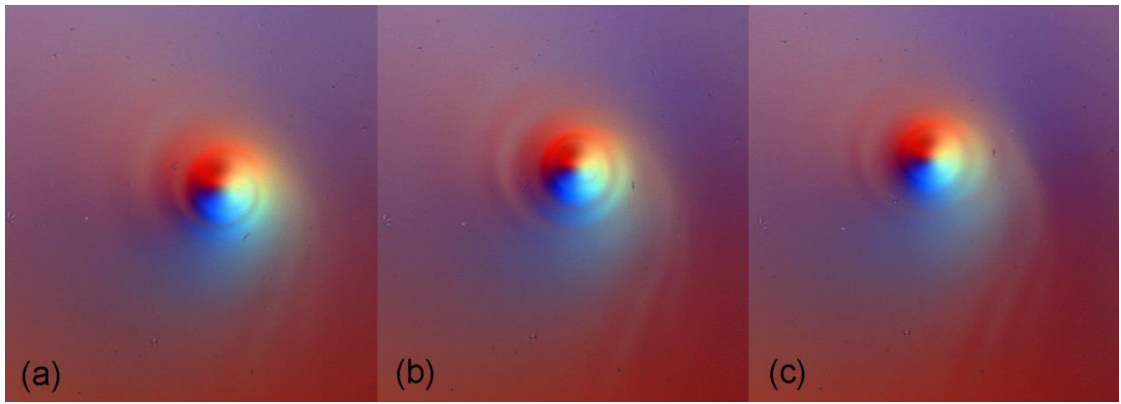

Figure 9. Sequence of the altimetric images of the vortex at $t = 10$ s (a), 12 s (b) and 14 s (c). Thin bands spiraling around the vortex are inertial waves.

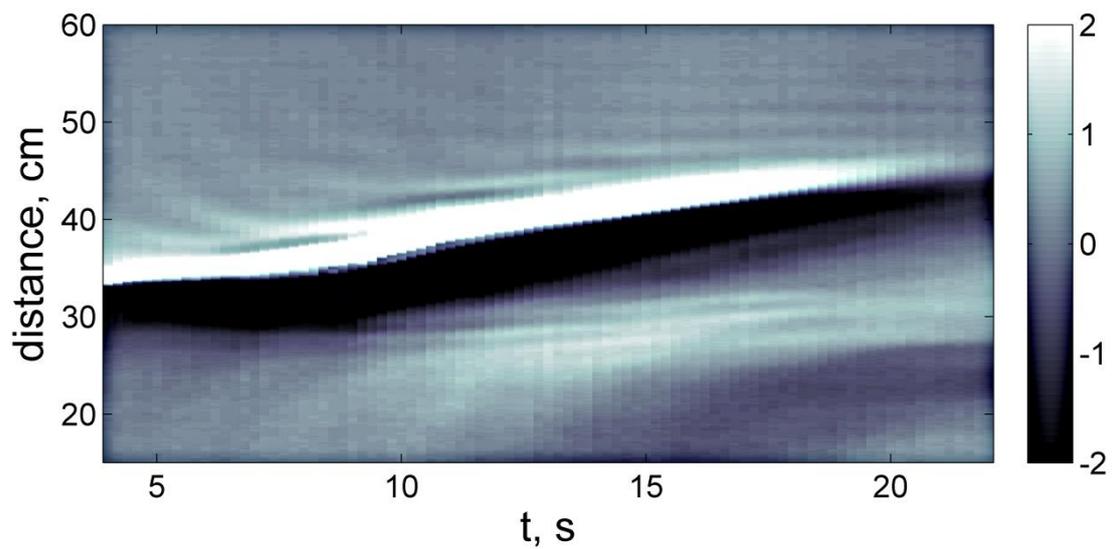

Figure 10. Hovmoeller (space-time) diagram of the geostrophic velocity measured along the straight line parallel to the vortex trajectory. Greyscale shows velocity in cm/s.

An insight into the dynamics of the Rossby wave radiation by a travelling vortex, can be gained by considering an energy spectrum of the flow in the wavenumber space. Two-dimensional energy spectrum is given by

$$E(k_x, k_y) = \frac{1}{2} |\mathbf{u}(k_x, k_y)|^2 , \qquad (26)$$

where ($k_x$, $k_y$) is the wavenumber and $\mathbf{u}(k_x, k_y)$ represents the discrete Fourier transform of the velocity vector. Figure 11 shows the evolution of the two-dimensional spectrum in the experiment 1 together with the spectra of the flow in our numerical simulations and in linear theory. All spectra have a typical inverted "S" shape. While in the simulations and theory the energy is mainly located in the lobes at the ends of the "S" shape, in the experiment, the significant energy is also concentrated at low $k_x$ that indicates that zonal modes are significant. To understand the observed spectra let us consider theoretical results by Lighthill[17] who described general properties of linear Rossby waves emitted by a moving disturbance (vortex). The disturbance moving with velocity $\mathbf{U}_t$ emits waves of frequency $\omega_0 + \mathbf{k} \cdot \mathbf{U}_t$, where $\omega_0$ is the natural frequency of the disturbance and $\mathbf{k} \cdot \mathbf{U}_t$ is the Doppler shift. For a steady disturbance, $\omega_0 = 0$ and the Doppler shift defines the wave radiation. A disturbance varying over a period of time $\Delta t$ emits transient waves with a spectrum of frequencies varying from 0 to, say, $10/\Delta t$. In our experiments the vortices are created by forcing over the time period $\Delta t$ such that the transients can be expected in the beginning of each experiment. After the forcing stops, the vortex evolves on a longer time scale determined by dissipation and by a loss of energy due to wave radiation. The wavevector $\mathbf{k}$ of a wave of particular frequency can then be determined from the dispersion relation

$$\omega_0 + \mathbf{k} \cdot \mathbf{U}_t = \frac{-k_x \beta}{k^2 + k_d^2}. \tag{27}$$

Black lines in Fig. 11 show $k_y$ as a function of $k_x$ calculated from Eq. (27) for a stationary disturbance, $\omega_0 = 0$. Instantaneous values of $\mathbf{U}_t$ were used to calculate the curves for the experiment while in the simulations and theory $\mathbf{U}_t$ was constant, $\mathbf{U}_t =$ 0.7 cm/s. While a close fit of the zero-frequency curves with the energy pattern in the theoretical spectrum is not surprising, the fit with the fully nonlinear numerical simulations is somewhat unexpected. The spectrum evolves from an approximately isotropic (Fig. 11 d) when the vortex is initially approximately axisymmetric, to the anisotropic spectrum (Fig. 11 e, f) which corresponds quite well to the linear dispersion relation (27). Transient waves with frequency corresponding to the forcing time $\Delta t$ can be expected in the experiments. In order to check where the transients are located in the wavenumber space and if their energy signature is noticeable in the experimental spectrum we plot the curves $\omega_0 = 2\pi/\Delta t$ in Fig. 11 a-c. At this frequency, Eq. (27) has two solutions; one is given by an (almost straight) line and another is a circle near the origin. However, there is no evidence of any concentration of energy along the line since the wavenumbers are relatively large there. The waves with large wavenumbers are not effectively radiated by the relatively large vortex as in our case. The low wavenumber waves corresponding to the circle near the origin can be radiated but they can hardly be distinguished from those corresponding to the zero-frequency curve.

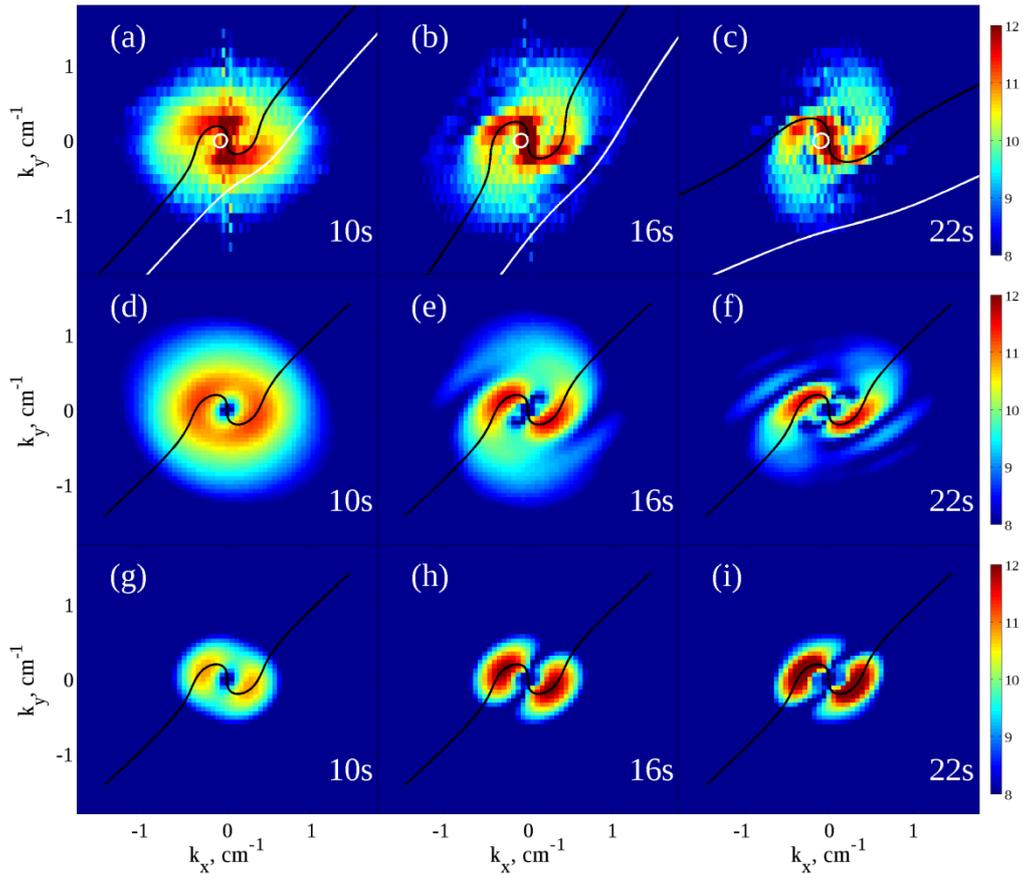

Figure 11. Two-dimensional energy spectra, $E(k_x, k_y)$ in experiment 1 (upper row, a - c), numerical simulations (middle row, d - f) and theory (bottom row, g - i) at $t = 10$ s, 16 s and 22 s. Color scale shows ln$E$. Solid black lines show the solution of Eq. (27) with $\omega_0 = 0$; The white curves show the solution of Eq. (27) with $\omega_0 = 2\pi/\Delta t$ where $\Delta t = 12$ s.

## VI. DISCUSSION

In this work, we have shown experimental evidence on the radiation of the Rossby waves by vortices moving on the $\beta$-plane. Vortices are self-propelled due to nonlinear interaction between primary monopolar flow field and the secondary dipolar flow which occurs due to $\beta$-effect as described by Reznik[27]. The measurements of the velocity due to the dipolar component of vortices demonstrated this effect. The cyclones generated by suction on our experiments are strongly nonlinear (similar to oceanic eddies[30]) and propagate to the northwest.

Travelling vortices radiate zero-frequency Rossby waves Doppler-shifted by $\boldsymbol{k} \cdot \boldsymbol{U_t}$ due to their motion. The pattern of waves is approximately parabolic such that the waves with relatively large wavenumber in (zonal) x-direction are to the East of the vortex and waves with small $k_x$ and approximately zonal crests are to the west as discussed by Rhines[33]. The radiation of the Rossby waves by vortices (eddies) can be one of the primary mechanisms of the creation of zonal jets observed in the oceans (Maximenko et al.[1, 2]). In fact, this mechanism is the basis of the important theoretical work by Rhines[34] on the dynamics of turbulence on the $\beta$-plane. In his original work, Rhines considered a field of closely packed eddies with a narrow spectrum around some wavenumber $k_0$ and assumed that $\beta$-term in the equation of motion is of the same order of magnitude as the nonlinear term. As a result, he obtained a wavenumber

$$k_\beta = (\beta/U_{rms})^{1/2} \tag{28}$$

which separates the eddies (turbulence) and waves in the spectral space. Here $U_{rms}$ is the root-mean-square fluid velocity at the energy containing wavenumber $k_0$. The Rhines scale has been widely discussed in the literature as a suitable measure of the meridional scale of the zonal jets. This work contributes to this discussion as follows. The energy spectra measured in our experiments as well as the spectra obtained in numerical simulations and theory suggest the scaling

$$k = (\beta/U_t)^{1/2}. \tag{29}$$

This expression gives the characteristic wavenumber for Rossby waves emitted by travelling vortices. Similar arguments as those applied for the Rhines scale can be used here to justify that this is an appropriate scaling for zonal flows/jets. Note that the translational velocity $U_t$ is used here. The translational velocity (at least for highly nonlinear vortices that are "self-driven" due to $\beta$-effect) can be much lower than the characteristic rotational velocity in the vortex (A >> 1). The characteristic rotational velocity can be interpreted here as an analogue of the $U_{rms}$, the velocity at the energy containing wavenumber $k_0$. For a field of closely packed vortices, originally considered by Rhines, there is no distinction between the two velocities because vortices are driven by strong interactions with each other such that their translational velocity is determined by the flow induced by their nearest neighbors rather than by the $\beta$-effect. However, one can imagine a field of more loosely packed vortices which only occasionally interact with each other but mostly driven by $\beta$-effect (and perhaps by the mean flow). This is, perhaps, the case in the ocean where mesoscale eddies are formed mostly at the eastern boundaries and then move westward across the oceans. In this case, the relatively subtle distinction between the velocities in Eqs (28) and (29) might be important.

**ACKNOWLEDGMENTS**

YDA gratefully acknowledges the support by the Natural Sciences and Engineering Research Council of Canada.